\documentclass[a4paper,10pt]{article}
\usepackage{epsf,latexsym,amsmath}
\usepackage{times}
\usepackage{amssymb}
\usepackage{graphicx}

\newcommand{\bpm}{\begin{pmatrix}}
\newcommand{\epm}{\end{pmatrix}}
\newcommand{\bmm}{\begin{matrix}}
\newcommand{\emm}{\end{matrix}}
\newcommand{\beq}{\begin{equation}}
\newcommand{\eeq}{\end{equation}}
\newcommand{\bc}{\begin{center}}
\newcommand{\ec}{\end{center}}

\begin{document}

\begin{center}

{\Large Braiding and entanglement in spin networks:}\\
{\Large a combinatorial approach to topological phases}
 
\vskip 1cm

{\large Z. K\'ad\'ar$^{\dag \,(1)}$,
 A Marzuoli$^{\ddagger \,(2)}$ 
and M Rasetti$^{\sharp\,(3)}$}

\vskip .5cm

{\small 
$^{\dag}$ Institute for Scientific Interchange Foundation, Villa Gualino, Viale Settimio Severo 75, 10131 Torino (Italy)\\
$^{\ddagger}$Dipartimento di Fisica Nucleare e Teorica, Universit\`a degli Studi di Pavia and Istituto 
Nazionale di Fisica Nucleare, Sezione di Pavia, via A. Bassi 6, 27100 Pavia (Italy)\\
$^{\sharp}$ Dipartimento di Fisica, Politecnico di Torino, corso Duca degli Abruzzi 24, 10129 Torino (Italy)}\\
\end{center}

\vskip 1cm

\begin{abstract}
\noindent The spin network quantum simulator relies 
on  the $su(2)$ representation ring
(or its $q$--deformed counterpart at $q=$ root of unity)
and its basic features naturally include (multipartite)
entanglement and braiding. 
In particular, 
$q$--deformed spin network automata are able to
perform efficiently approximate calculations
of topological invarians of 
knots and $3$--manifolds.
The same algebraic background is shared 
by $2D$ lattice models 
supporting topological phases of matter that have recently 
gained much interest in condensed matter physics. 
These developments are motivated by the possibility 
to store quantum information fault--tolerantly 
in a physical system supporting fractional statistics
since a part of the associated Hilbert space is insensitive 
to local perturbations.
Most of currently addressed approaches are framed within 
a `double' quantum Chern--Simons field theory, whose quantum amplitudes
represent  evolution histories of local lattice degrees of freedom.
 
We propose here a novel combinatorial approach based
on `state sum' models of the Turaev--Viro type associated with 
$SU(2)_q$--colored triangulations of the ambient $3$--manifolds.
We argue that boundary $2D$ lattice models (as well as observables in the form of colored graphs
satisfying braiding relations) could be consistently addressed.
This is supported by the proof that 
the Hamiltonian of the Levin--Wen condensed string net model in a surface 
$\Sigma$ coincides with the corresponding Turaev--Viro amplitude on $\Sigma \times [0,1]$ 
presented in the last section. 
\end{abstract}

\vspace{15pt}


\vspace{10pt}


\vspace{40pt}

\noindent
---------------------------------------\\
(1) kadar@isi.it\\
(2) annalisa.marzuoli@pv.infn.it\\
(3) mario.rasetti@polito.it\\

\vfill
\newpage

\section{Braiding and entanglement in spin networks}

By {\em spin networks} we mean here 
`computational' graphs the nodes and edges of
which are labeled by dimensions of $SU(2)$ irreducible 
representations (irreps) and by $SU(2)$ recoupling transformations, respectively. For this reason 
spin networks can be thought of as a generalized quantum computational framework 
for dealing with unitary transformations 
among (entangled) many--angular momenta states.
Actually, the computational space of the {\em spin network simulator}\cite{MaRa} 
encodes in its very definition the representation ring
of the Lie group $SU(2)$ given by finite--dimensional Hilbert spaces
supporting irreps of $SU(2)$ endowed 
with  two  operations, tensor product  $\otimes$ 
and direct sum $\oplus$ (providing a ring 
structure over the field $\mathbb{C}$) as well as
unitary operators relating (multiple tensor products of) 
such spaces.
 
Unlike the usual standard
circuit model\cite{NiCh}, here it is possible to handle  eigenstates
of $N$  (pairwise coupled) angular momenta labeled by
integers and half--integers $j_1,j_2,\ldots,j_N$ (in $\hbar$ units) and
not simply $N$--qubit states for which
 $j_1=j_2= \ldots =j_N=1/2$. The (re)coupling theory of $N$ $SU(2)$
 angular momenta provides the whole class of unitary transformations that can be 
 performed on many--body quantum systems described by this kind of states\cite{BiLo9}.
 Any such general transformation --expressed in terms of a $3nj$--coefficient
($N=n+1$) when a choice of basis sets is made
explicit-- can be split into a finite sequence of
two basic unitaries, called {\em twist} (or `trivial braiding')
and {\em associator} in the language of tensor categories\cite{FuSc}. 
The former acts on the (ordered) tensor product
of two Hilbert spaces $V, W$ supporting irreps of $SU(2)$ by swapping them, namely
\begin{equation}\label{twist}
T_{V,W}\,:\;V\,\otimes\,W\;\rightarrow\;W\,\otimes\,V\;\;\;
\text{with}\;\;\;T_{W,V} \circ T_{V, W}\,=\text{Id}_{V \otimes W}
\end{equation} 
and its action 
on a binary coupled state amounts to a trivial phase transform.\\
The associator $F$ relates different binary bracketing structures
in the triple tensor product of irreps $V,U,W$ (intertwiner)
 \begin{equation}\label{assoc}
F\,:\;(V\,\otimes\,U)\,\otimes\,W\;\rightarrow\;V\,\otimes\,(U\,\otimes\,W)
\end{equation}
and is implemented on any binary coupled state as a transformation involving one Racah--Wigner
$6j$ symbol.
Notice that both \eqref{twist} and
\eqref{assoc} are isomorphisms  but the associator reflects a physically
measurable modification of the way in which intertwiner spaces are coupled.
More complicated multiple
tensor products can be related by various combinations of 
braidings and associators, so that the basic isomorphisms must satisfy
compatibility conditions, a so--called pentagon identity
and two hexagon identies\cite{FuSc} (in $SU(2)$ angular momentum theory
they correspond to the Biedenharn--Elliott identity and 
the Racah identity, respectively\cite{BiLo9}). 

The most effective way of dealing 
with non--trivial braiding operators --to be used in connection
with the study of both braid group representations and braid statistics 
(typically occuring in anyonic systems)-- is to move 
to the representation ring $\mathfrak{R}$ $(SU(2)_q\,)$ (modular tensor category)
of the $q$--deformed 
Hopf algebra of the Lie group $SU(2)$, $SU(2)_q$ ($q$ a root of unity).
Then the operation $T$ in \eqref{twist} turns out to be substituted
by a non--trivial braiding
\begin{equation}\label{braidmor}
R_{V,W}\,:\;V\,\otimes\,W\;\rightarrow\;W\,\otimes\,V \;\;\;
 \text{with}\;\; R_{W,V} \circ R_{V, W} \neq
\text{Id}_{V \otimes W}. 
\end{equation}
The $q$--{\em deformed} spin network model of 
computation\cite{GaMaRa1} is modelled on the $q$--representation ring
\begin{equation}\label{qcategory}
(\,\mathfrak{R} (SU(2)_q\,)\,;R\,;F\,),
\end{equation}
where $F$ denotes from now on the $q$--counterpart
of the associator \eqref{assoc}. Once suitable basis sets are
chosen, the isomorphisms $R$ and $F$ can be made
explicit. In particular, $F$ contains the 
$q$--deformed counterpart of the $6j$--symbol and, regarding it 
as a unitary matrix, it is also referred to as `duality' (or
`fusion') matrix in conformal field theories\cite{GoRuSi}.

The framework outlined above was exploited
in a series of papers\cite{GaMaRa1,GaMaRa2}, where families of 
$q$--deformed spin network automata were implemented
for processing efficiently classes of computationally--hard problems 
in geometric topology --in particular, approximate calculations
of topological invarians of links\cite{Jon} (collections of 
knots) and $3$--manifold\cite{ReTu}. A prominent role was played there by
`universal' unitary braiding operators associated with 
representations of the braid group in $\mathfrak{R}$ $(SU(2)_q\,)$. 
Suitable traces of matrices
of these representations provide polynomial
invariants of $SU(2)_q$--colored links, while 
weighted sums of the latter give topological
invariants of $3$--manifolds presented as complements 
of  knots in the $3$--sphere.  These  invariants 
are  in turn recognized as partition functions and vacuum 
expectation values of physical observables in $3$--dimensional 
Chern--Simons--Witten (CSW)  Topological Quantum Field Theory (TQFT)\cite{Wit1}.
The CSW environment actually provides not only a physical interpretation of such
quantities but it is universal in the sense that includes,
besides the quantum group interpretation quoted above,
monodromy representations of the braid group arising in
a variety of (boundary) conformal field theories\cite{GoRuSi} 
(where point--like excitations confined in $2$--dimensional
regions evolve along braided worldlines).

It is worth mentioning that the $q$--spin network 
approach represents  a naturally `discretized' version of the topological
framework for quantum computation proposed a few years ago
and further improved recently\cite{Fre}. In the light of some
basic questions raised by that paper, we are going to
outline a novel approach to the whole matter of topogical phases (section 2)
providing in particular a transparent combinatorial description
of condensed strings nets (section 3).

\section{$SU(2)_q$--colored  triangulations: $3D$ invariant partition functions and
induced $2D$ lattice models}

The issue of connections between (topological) gauge theories
in $3$ spacetime dimensions and $2D$
(integrable) lattice models\cite{Wit2} 
has been intensively investigated over the years
in a variety of different contexts. 
The renewed interest driven by the search for a fault--tolerant
quantum computer  based on manipulations of Non--Abelian quantum Hall
states\cite{Fre,toricc} represents a major challenge on both
experimental and theoretical grounds.
Most of the currently addressed approaches are framed within
a CSW environment, but 
an {\em ab initio} discretized framework
could reveal much more effective, as briefly 
outlined in the following.

The Turaev--Viro (TV) `state sum' model\cite{tv}   
provides   a well--defined ({\em i.e.} finite)
topological invariant for any closed $3$--manifolds $M^3$. 
Given a triangulation $T^3$ of $M^3$, namely a dissection into tetraheda,
 an $SU(2)_q$--coloring is assigned to it
according to a set of admissible initial data (not made explicit here).
A state functional for $T^3$ is  built, where the $N_0$ vertices, 
$N_1$ edges and $N_3$ 
tetrahedra are suitably weighted. Finally, a summation over all
admissible colored triangulations of $M^3$ 
is performed, the resulting functional being invariant under
a set of combinatorial moves ensuring that its value depends
only on the topological type of the manifold $M^3$. The TV state sum reads
\begin{equation}\label{TVstsum}
\mathbf{Z}_{\,TV}\,[M^3;q]\,=\,\sum_{\{j\}}\;\mathbf{w}^{-N_0}\,
\prod_{A=1}^{N_1} \mathbf{w}_A
\,\prod_{B=1}^{N_3} \;
\begin{Bmatrix}
j_1 & j_2 & j_3 \\
j_4 & j_5 & j_6
\end{Bmatrix}_B \,,  
\end{equation}
where the summation is over all colourings $\{j\}$ labeling irreps of $SU(2)_q$
\newline 
($q=\exp\{2\pi i /r\}$, with $\{j=0,1/2,1 \dots, r-1\}$); $\mathbf{w}_A\doteq$ 
$(-1)^{2j_A}[2j_A+1]_q$ 
where $[\,]_q$ denote the quantum dimension of the irrep; 
$\mathbf{w}=2r/(q-q^{-1})^2$ and $\{\,:::\}_B$ represents the $q-6j$ symbol 
whose entries are associated with the six edges of tetrahedron $B$.

The invariant \eqref{TVstsum} equals the 
square modulus of the Reshetikhin--Turaev invariant\cite{ReTu}, which in turn represents
the Chern--Simons partition function $\mathbf{Z}_{\,CS}$ for an {\em oriented} $3$--manifold
$\mathcal{M}^3$, namely 
$\mathbf{Z}_{\,TV}\,[\mathcal{M}^3;q\,]\,\longleftrightarrow$
$|\,\mathbf{Z}_{\,CS}\,[\mathcal{M}^3;k\,]\,|^2\,,
$
where $k= 2(r-1)$ is the level of Chern--Simons functional.
This feature is crucial in view of applications
to topological phases, where  a `doubling' 
of the basic CS setting is needed\cite{Fre}.

The TV model can be further extended to deal with both
$3$--manifolds with $2D$ boundary component(s) and 
`observables' related to embedded links and (ribbon) graphs.
The extension to a pair $(M^3, \Sigma)$ with $\Sigma \equiv \partial 
M^3$ can be carried out in two different ways,
depending on whether the boundary surface $\Sigma$ 
inherits a fixed triangulation\cite{ks}
or a `fluctuating' one\cite{CaMa}.
Observables in the form of colored graphs, 
satisfying braiding relations, can be consistently 
introduced\cite{ks,db} in any oriented 
triangulated compact manifold $(M^3, \Sigma)$.

On the basis of such results we are currently addressing
a combinatorial reformulation of  theoretical
foundations of topological phases\cite{KaMaRa}.
A first result is discussed in the last section.

\section{Topological spin liquids: a combinatorial description}

We will now briefly introduce the condensed string nets of Levin and Wen\cite{levinwen} in $(2+1)$ dimension and
show that the projector to the ground state of the boundary model can be obtained from the corresponding Turaev--Viro 
partition function $Z_{TV}(\Sigma\times [0,1],X)$ for the cylinder $M^3 = \Sigma\times [0,1]$
($\Sigma$ is oriented and $X$ denotes a fixed
identical triangulation on both $\Sigma\times\{0\}$ and $\Sigma\times\{1\}$). 

Consider the string net model defined on the honeycomb lattice\cite{toricc} 
with microscopic degrees of freedom associated to 
the edges. The degrees of freedom are called string types and are endowed
with  $N+1$ labels and orientations. The change of 
orientation of an edge is equivalent to the change of the label $i$ to its dual $i*$ and the $0$ label is 
self--dual. The Hamiltonian is a sum of potential and kinetic energy. When the relative coefficient of the kinetic 
piece is large, the ground state is a condensate of a dense string net. The universal long distance behaviour of these
nets is characterized by the so--called fixed--point wave functions on the space of all configurations. The 
fixed--point wave functions $\Phi$ are in one--to--one correspondence with modular tensor categories \cite{Tur} 
and obey the following rules (written by using the original notation\cite{levinwen}, 
although slightly different from the conventions in \eqref{TVstsum}):
\begin{align}
 \Phi
\bpm \includegraphics[height=0.3in]{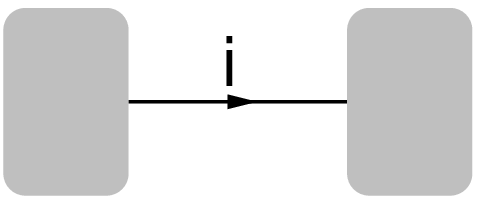} \epm  =&
\Phi 
\bpm \includegraphics[height=0.3in]{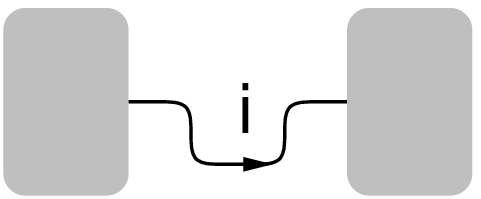} \epm
\label{topinv}
\\
 \Phi
\bpm \includegraphics[height=0.3in]{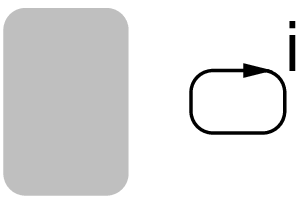} \epm  =&
d_i\Phi 
\bpm \includegraphics[height=0.3in]{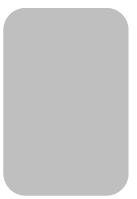} \epm
\label{clsdst}
\\
 \Phi
\bpm \includegraphics[height=0.3in]{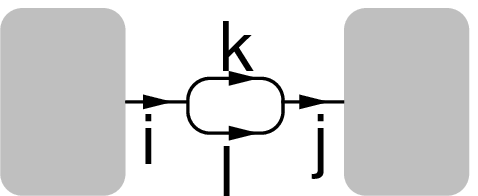} \epm  =&
\delta_{ij}
\Phi 
\bpm \includegraphics[height=0.3in]{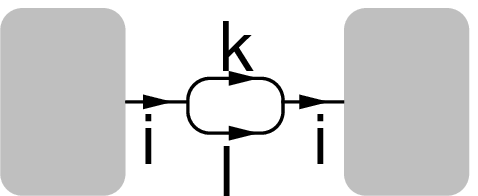} \epm
\label{bubble}\\
 \Phi
\bpm \includegraphics[height=0.3in]{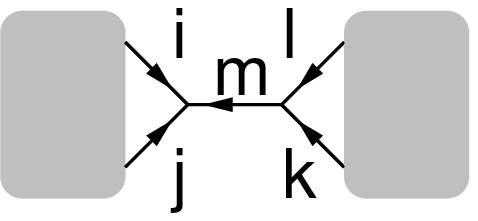} \epm  =&
\sum_{n} 
F^{ijm}_{kln}
\Phi 
\bpm \includegraphics[height=0.28in]{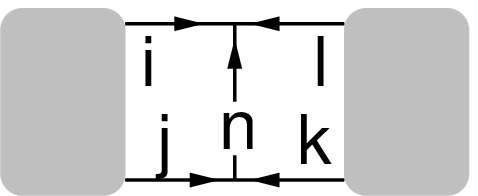} \epm 
\label{fusion} 
\end{align}

\noindent where the $F$ matrix and the quantum dimensions $d_i$ satisfy consistency conditions
given by    

\begin{eqnarray}
F^{ijk}_{j^*i^*0} &=& \frac{v_{k}}{v_{i}v_{j}} \delta_{ijk} \nonumber \\ 
F^{ijm}_{kln} = F^{lkm^*}_{jin} &=& F^{jim}_{lkn^*} = F^{imj}_{k^*nl}
\frac{v_{m}v_{n}}{v_{j}v_{l}}
\nonumber \\
\sum_{n=0}^N 
F^{mlq}_{kp^*n} F^{jip}_{mns^*} F^{js^*n}_{lkr^*}
&=&
F^{jip}_{q^*kr^*} F^{riq^*}_{mls^*} 
\label{pent}
\end{eqnarray}
with $v_i=\sqrt{d_i}$ ($={\bf w}_i$ in TV notation) and $\delta_{ijk}=0,1$. 
The latter is the branching rule: the triple $\{ijk\}$ is 
admissible at a vertex when $\delta_{ijk}=1$. These rules coincide with those of Turaev and Viro\cite{tv} for 
the evaluation of closed manifold invariants in three dimensions and the (first two lines of the) consistency 
conditions encode the standard properties of the $q-6j$ symbol suitable normalized as 
\beq 
d_n \left\{\begin{array}{lll}i&j&m\\k&l&n\end{array}\right\}=F^{ijm}_{kln} ,
\label{norm}
\eeq
while the last relation in (\ref{pent}) is the Biedenharn--Elliott (pentagon) identiy ({\em cfr.}
section 1).

The Hamiltonian for the string--net model on the honeycomb lattice reads
\beq H=-\sum_I Q_I-\sum_p B_p\ , \eeq
where $I$ runs over vertices and $p$ over plaquettes. 
$Q_I=\delta_{ijk}$ with $i,j,k$ being the colours of the edges 
incident to the vertex $I$. The `magnetic constraints' are given by the sum $\sum_{s=0}^N a_s B_p^s$ with $B_p^s$ to 
be described below. We consider the case when the coefficients are
\beq a_s=\frac{d_s}{\sum_{i=0}^N d_i^2}\ .\label{as}\eeq
This normalization prescription corresponds to the existence of smooth 
continuum limit of the lattice model in one hand and gives the precise 
weight of the Turaev--Viro evaluation on the other, as proved  below.

\vskip .5 cm

\subsection*{The magnetic constraints}

There is a nice intuitive way of determining the matrix elements of the magnetic constraints described in 
appendix C\cite{levinwen}. In short, for each hexagon
of the lattice, one draws a smaller concentric hexagon with labels $s$ at every edge 
in hexagon $p$, then {\bf (i)} connects the hexagons $p$ and $s$ with new edges 
and labels them with $0$ (this does not change 
$\Phi$) and {\bf(ii)} use the rules (\ref{fusion}) repeatedly until one reaches the single hexagon again. 
The resulting coefficients define the matrix elements of the constraint given esplicitly by
\beq\begin{array}{l}
B^s_{p}\Big|\bmm\includegraphics[height=0.8in]{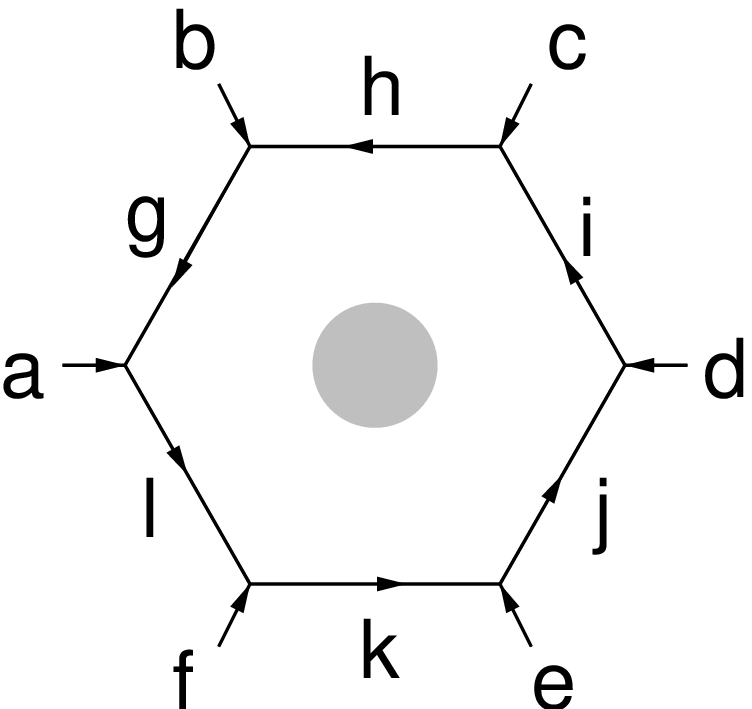}\emm\Big\rangle=\\
{\displaystyle\sum_{g'h'i'j'k'l'}
F^{bg^*h}_{s^*h'g'^*}
F^{ch^*i}_{s^*i'h'^*}
F^{di^*j}_{s^*j'i'^*}
F^{ej^*k}_{s^*k'j'^*}
F^{fk^*l}_{s^*l'k'^*}
F^{al^*g}_{s^*g'l'^*}\Big|}
\bmm\includegraphics[height=0.8in]{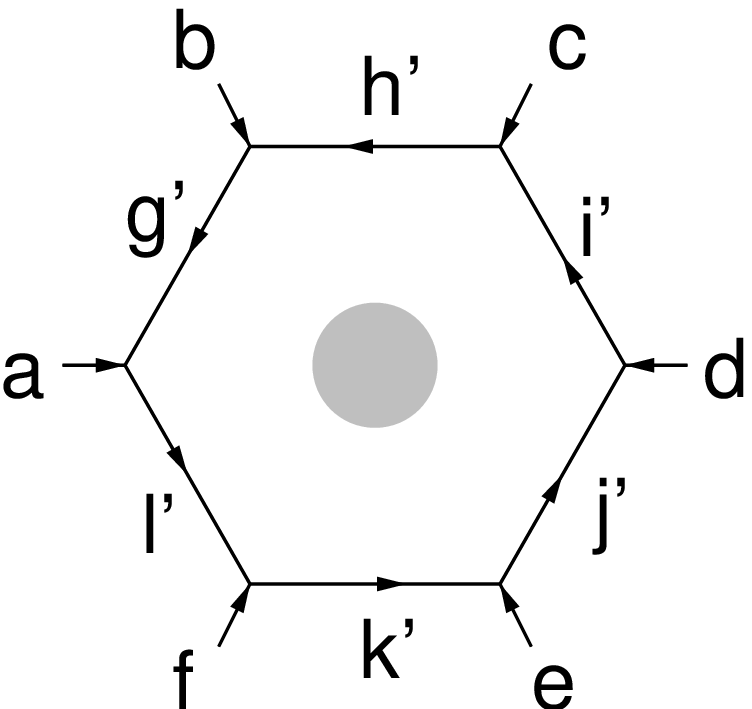}\emm\Big\rangle
\end{array}\eeq
The operators $B_p$ with the coefficients given by (\ref{as}) are projectors (as the $Q_I$'s) and all these constraint
operators commute with each other.

\begin{figure}
\bc\includegraphics[width=3cm]{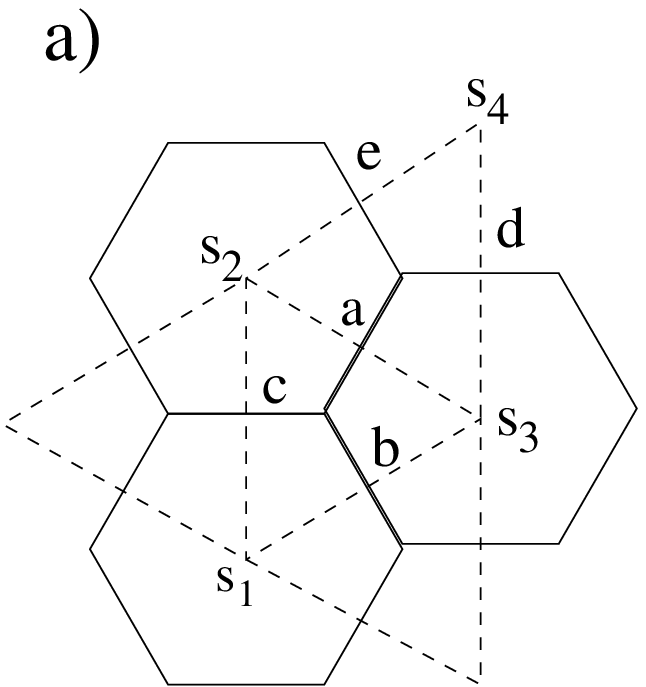}
\includegraphics[width=7cm]{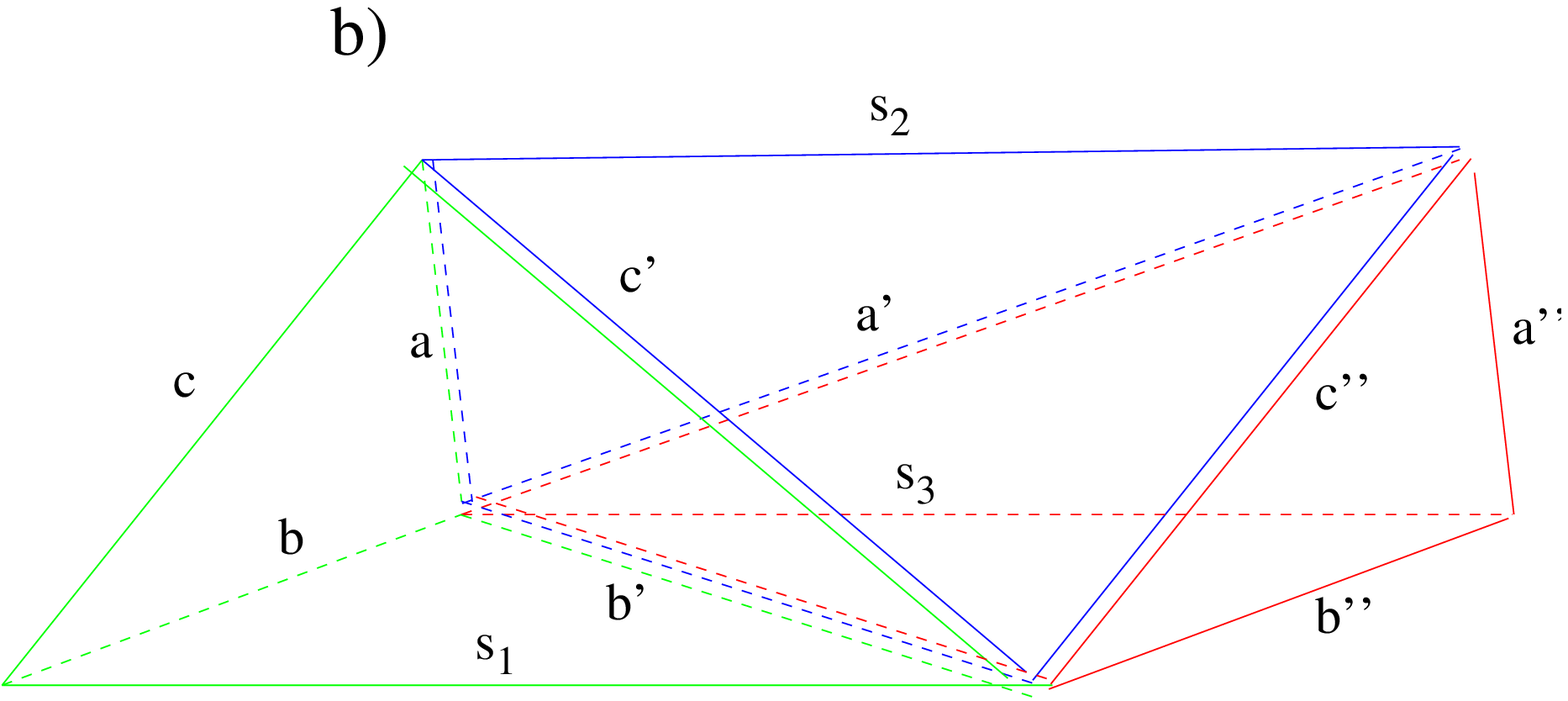}
\caption{\label{dgp} a) shows a part of the honeycomb lattice and its dual graph. In b) the `plane' of the
triangle $abc$ is the `plane' of the honeycomb lattice and there is another one with triangle 
$a"b"c"$. The matrix element of the projector is interpreted as a three dimensional extrapolation between the bra 
and the ket string net state given by a triangulation. The prism in the figure correspond to a part of the projector 
$P$ where there is a tetrahedron for every associated $F$ symbol.}
\ec\end{figure}

The image of the projector $P=\prod_q Q_q\prod_p B_p$ is the ground state of the model. We 
are going to  
show explicitly that
\beq P=Z_{TV}(\Sigma\times [0,1],X)\ ,\label{res} \eeq
where $\Sigma$ is the (oriented) surface on which the honeycomb lattice is defined 
and X is the dual triangulation of 
$\Sigma\times\{0\}\simeq\Sigma\times\{1\}\simeq\Sigma$.
To build the dual graph one places a vertex in the center 
of each face and connects the new vertices in neighbouring faces. 
In particular, the dual of the trivalent 
honeycomb lattice is a triangular graph as shown in Fig.\ \ref{dgp}a).

Let us start constructing $\prod_p B_p$. Each operator $B_p^s$ contains six $F$ symbols corresponding to the six 
vertices of a hexagon. We collect the three $F$ symbols from three different $B_p^s$ operators corresponding to a 
given vertex. Their order does not matter since the $B$ operators commute. Choosing the order 
$B_{p_1}^{s_1}B_{p_2}^{s_2}B_{p_3}^{s_3}$ ({\em cfr.}  Fig.\ \ref{dgp}a) we 
get the contribution $F_{s_1 b' c'}^{acb} 
F_{s_2 c" a'}^{b'ac'} F_{s_3 a" b"}^{c"b'a'}$. Note also that we have omitted denoting the 
orientation for the sake of simplicity, but it is not difficult to incorporate it consistently. Comparing this 
expression with Fig.\ \ref{dgp}b), we find that it corresponds to the triangulation of the prism with boundary 
triangles $abc$ and $a"b"c"$. In this identification we have associated a $q-6j$ or $F$ symbol to every tetrahedron, 
which is the prescription of the Turaev--Viro state sum (\ref{TVstsum}). 

The second step consists in checking whether the gluing along the 
triangulated boundary quadrilaterals can be made consistent with the algebraic pattern. Using the freedom of 
multiplying the F symbols at a given vertex in arbitrary order (due to the commuting $B_p$ operators that they are 
part of), the consistency can be achieved. It is clear now that we have a triangulation given by the dual graph with 
edge labels $a"b"c"d"\dots$ and an identical one with labels $abcd\dots$ and these bound a $3$--dimensional 
triangulation given by triangular prisms suitably glued together. 

We now check whether the weights 
associated to edges and vertices match that of the TV prescription. 
Taking into account the normalisation (\ref{norm}) we find that there is a factor of $d_i$ associated to each internal 
edge denoted by $a'b'c'd'\dots$. Because of the coefficient (\ref{as}) it is also true for the internal edges 
`perpendicular' to $\Sigma$ labelled by $s_i$ in the figures (this weight comes from the numerator of $a_s$, 
the denominator 
will be associated to vertices bounding these edges). To the boundary edges with labels $a"b"c"d"\dots$ there is
also a $d_{a"},d_{b"}$ etc. associated, whereas to those with label $abcd\dots$ there is no non--trivial weight. 
This is consistent with the TV prescription\footnote{For example we can introduce a black and white colour on 
boundaries and glue only black to white boundary, that is one with $d_i$ weights to one with no weights of vertices.}. 
The last type
of simplexes we have and not discussed yet are the boundary vertices. We associate the square root of the 
denominator $(\sum_i d_i^2)^{-1/2}\equiv D^{-1}$
($={\bf w}^{-1}$ in TV notation) of the coefficient $a_s$ to the two boundary vertices connected by 
the edge labeled by $s$. This way we covered both boundaries and found agreement with the TV prescription.

The last step is the sum over internal colouring, which has to be performed to get the partition function. 
This corresponds to the summation over the labels $s_i$ and $a'b'c'd'\dots$, which 
come from $\sum_i a_s B_p^s$ and the matrix
multiplication in $\prod_p B_p$. Finally, note, that the electric 
constraints $\prod_q Q_q$ are taken care of by the product $\prod B_p$ as had 
non--admissible labels met at a vertex, a corresponding $q-6j$ symbol gives 
zero. This concludes the proof of (\ref{res}).

\end{document}